\documentstyle[12pt]{article}
\begin{document}
\title{ Expansion in the Width and Collective Dynamics of a Domain Wall }
\author{by \\
\\
H. Arod\'z   \\
        \\
Institute of Physics, Jagellonian University,
Cracow \thanks{Address: Reymonta 4, 30-059 Cracow, Poland.}
        \thanks{E-mail: ufarodz@thrisc.if.uj.edu.pl} }
\date{ $\;\;\;$}
\maketitle

\thispagestyle{empty}

We show that collective dynamics of a curved domain wall in a
(3+1)-dimensional relativistic scalar field model is represented by  
Nambu-Goto membrane and (2+1)-dimensional scalar fields defined on the 
worldsheet of the membrane. Our argument is based on a recently proposed 
by us version of the expansion in the width. Derivation of the expansion 
is significantly reformulated for the present purpose. Third and fourth 
order corrections to the domain wall solution are considered. We also 
derive an equation of motion for the core of the domain wall. Without 
the (2+1)-dimensional scalar fields this equation would be nonlocal. \\
$\;\;$   \\
March 1997        \\
TPJU-2/97      \\
\pagebreak

\setcounter{page}{1}

\section{Introduction}

One of the  most interesting branches of field theory is devoted to 
spatially extended topological solitons in 3+1 dimensional
space-time such as domain walls or vortices. By now it is clear that 
these solitons, mathematically represented by certain nontrivial 
solutions of classical field equations, play an important role in many 
phenomena in condensed matter physics \cite{1,2,3}, particle physics 
\cite{4}, and perhaps also in cosmology \cite{5}.
Dynamics of the extended solitons is extremely rich, but also it poses
rather hard problems because of its strongly nonlinear character. In
contradistinction to the case of solitons in one-dimensional space, there
are no methods of constructing  generic exact domain wall or vortex
solutions. Till now most theoretical results have been obtained with the
help of numerical methods, see, e.g. \cite{6,7,8}, eventhough there are
certain remarkable exact analytical results like the discovery of
travelling waves on a straightlinear vortex
\cite{9}, or of the $90^{o}$ scattering of straightlinear parallel
vortices \cite{10}.

Analytical approaches to evolution of a generic curved vortex/domain wall 
are scarce. In most of them one attempts to reduce the dynamics to much
simpler dynamics of a string/membrane.  This line of thought has been
initiated in the seminal paper \cite{11},
continued in \cite{12,13} and later on in
[14--21]. It has been found already in \cite{11,12} that in the
case of local vortices, e.g., in the Abelian Higgs model, one may
approximately describe motion of the vortex using the simple classical
Nambu-Goto string model as the leading approximation. Dynamics of internal
degrees of freedom of the vortex is completely neglected in this
approximation. In a reference frame co-moving with the vortex all fields
have by assumption  the same form as for a static,
straightlinear vortex. In other words,
the field structure of the vortex in the co-moving reference frame is kept
frozen. The case of domain walls is strictly analogous --- here one has the
Nambu-Goto membrane instead of the string.

In the case of a global vortex a long range field is present. In the 
leading approximation the collective dynamics  of the vortex is described
in terms of the Nambu-Goto string coupled to a massless 2-form
gauge field \cite{22}.

Our present paper is devoted to the problem how to compute corrections
to the collective dynamics which are due to the internal dynamics of the
vortex/domain wall. Specifically, we further
develop the approach proposed in \cite{23}. In the literature one
can also find  other attempts to calculate  evolution of the curved
vortex/domain wall beyond the leading approximation.  Let us review
them briefly.

In the most popular approach [14--21] the
string/membrane is identified with the locus of zeros of a scalar field,
that is with the core. Influence of the internal dynamics
on the collective motion is taken into account by
corrections to the Nambu-Goto  action for the core. The total action
defines the so called {\it effective classical string/membrane
model}. It turns out that such effective models are rather unpleasant.
Equations of motion contain higher derivatives with respect to
time and consequently they have unphysical runaway type
solutions. Moreover, it seems that the effective models
should be nonlocal, see \cite{24} where caveats concerning this approach
are presented and Section 4 of the present paper.

In another approach \cite{25} one derives directly from field
equations approximate equations of motion for  certain
averages, e.g. for an average radius in the case of a ring-like
vortex, in an analogy with the Ehrenfest's approach to
evolution of a quantum mechanical wave packet in an external potential.
It is clear that the knowledge of time evolution of such averages is not
sufficient in order to recover the time evolution of the fields of
the vortex/domain wall.

In the polynomial approximation  \cite{26} the string/membrane
is identified with the core, but it gives only a
part of the collective dynamics. The other collective degrees of
freedom are represented by coefficients of certain polynomials which
approximate the fields inside the vortex/domain wall. They can be
regarded as scalar fields defined on the worldsheet of the
string/membrane. Here one avoids shortcomings of the previous
two approaches. On the other hand, the
accuracy of this approximation has been checked only in the simplest cases
of static straightlinear vortices and planar domain walls. Moreover,
it is likely that this approach should be generalised to include
radiation of massive fields from the extended solitons. Such radiation
could dump oscillations of the width which in the present formulation
of the method can persist indefinitely (this problem is currently
under investigation).

The approach proposed in \cite{23} is based on the expansion
in the width for the vortex/domain wall solution of the field equations.
It is constructed along the lines of Hilbert-Chapman-Enskog method
for singularly perturbed evolution equations \cite{27}. In this approach
certain consistency conditions play a crucial role. Co-moving
string/membrane is of the Nambu-Goto type in all orders, but it does not 
coincide with the core. In addition to the string/membrane there are also 
scalar fields defined on the worldsheet of the co-moving string/membrane. 
We avoid the unpleasant effective classical models with higher derivatives.
In contradistinction to the case of polynomial approximation we have
here physically meaningful expansion parameter, that is the width
defined as the inverse mass ($M^{-1}$) of the scalar field. More
precisely, the expansion parameters are given by dimensionless
ratios of the width to local curvature radia of the string/membrane
worldsheet. In the case of a domain wall the terms in the
expansion have been calculated up to the order $M^{-2}$.
These results have been compared with  purely numerical solutions for
cylindrical and spherical domain walls \cite{28} --- it has turned
out that the approximate analytical solution is astonishingly accurate.

Expansion in the width was invoked also in the earlier papers, e.g.,
in \cite{14,15,17,21} where a version of it was used on intermediate
stages of derivations of the effective string/membrane models.
However, in those papers the consistency conditions were not taken
into account. This is a crucial difference because in our version of
the expansion just these conditions make the calculation of the
effective classical string/membrane action superfluous.
More detailed comparison of our version of the expansion with the other
ones can be found in the first of papers \cite{23}. See also Section 4
of the present paper.

The construction of the expansion given in \cite{23} has two main
elements: a special co-moving coordinate system \cite{12} introduced
to maintain the Lorentz invariance order by order in the expansion, and
the consistency conditions. While the use of the co-moving
coordinates could be regarded as a rather elegant
and relatively simple way to secure the Lorentz
invariance, the consistency conditions appeared in  \cite{23} as formal,
purely mathematical observations. In spite of the crucial role they
played in determining the vortex/domain wall solution, their
field theoretical meaning was hidden. Moreover, the calculations were rather
cumbersome and even for the domain wall, which is simpler than the vortex,
we were able to obtain only the first and the second order corrections.
We found  that the core did not coincide with the co-moving
string/membrane, and that the co-moving string/membrane obeyed Nambu-Goto
equation in all orders, but we did not derive any equation of motion 
for the core. Such
equation would be particularly useful for comparison with the effective
string/membrane approach in which the string/membrane is
identified with the core.

In the present paper we concentrate on the case of a domain wall. 
Analogous results can be obtained also for a vortex but the
necessary calculations are much longer. We reformulate the derivation
of the expansion in such a way that the (2+1)-dimensional scalar fields
defined on the worldsheet of the co-moving membrane, which have been 
introduced in \cite{23} for purely mathematical reasons, acquire a physical
interpretation --- they are identified with perturbative contributions
to the values of the original scalar field on the membrane.
We also find a field theoretical interpretation of the consistency
conditions: they coincide with certain Euler-Lagrange equation. Moreover,
derivations of the higher order terms in the expansion are
significantly simpler in the new formulation.  Now we can consider
corrections in the next two orders, that is in the $M^{-3}$ and
$M^{-4}$ orders. Order by order in the $1/M$ expansion we 
express profile of the domain wall by the collective degrees of
freedom given by the co-moving membrane and the (2+1)-dimensional scalar
fields. Finally, we derive an approximate equation of
motion for the core. It
turns out to be different from equations obtained in the effective action
approaches. In particular, it does not contain higher derivatives.

The main conclusion of our paper is that the collective dynamics of the
domain wall beyond the leading approximation involves the (2+1)-dimensional
scalar fields defined on the worldsheet of the Nambu-Goto membrane.
Elimination of those fields would lead to a nonlocal membrane model for the
collective dynamics.

The plan of our paper is as follows. In Section 2 we present
Euler-Lagrange equations and we discuss certain important consequences of
them. In Section 3 we consider the extrinsic curvature corrections to the
domain wall solution up to the $M^{-4}$ order. The equation of motion for
the core is derived in Section 4.  Section 5 is devoted to
a discussion of our results. In Appendix A we have collected formulas
relevant for the transformation to the co-moving coordinates. In Appendix B
we present solutions of certain linear ordinary differential equation.
In Appendix C we list functions and constants which appear in Section 3.

\section{The consistency conditions as Euler --
Lagrange equation }

We consider a domain wall in the model defined by the following Lagrangian 
\begin{equation}
{\cal L} = -\frac{1}{2}\eta_{\mu\nu}\partial^{\mu}\Phi \partial^{\nu}\Phi
-\frac{\lambda}{2}(\Phi^{2}-\frac{M^2}{4\lambda})^2, \label{1}
\end{equation}
where $\Phi$ is a single real scalar field, $(\eta_{\mu\nu})$=diag(-1,1,1,1)
is the space-time metric, and $\lambda,M$ are positive constants.
Two vacuum values of $\Phi$ are equal to $\pm M/2\sqrt{\lambda}$.
The mass of the corresponding scalar particle is equal to $M$.

The domain wall appears when a solution  $\Phi$ of the Euler-Lagrange
equation obtained from the Lagrangian (\ref{1}) smoothly
interpolates between the two vacuum values. Then, at each
instant of time the field $\Phi$ vanishes
somewhere in the interior of the domain wall. The locus of these zeros is
assumed to be a smooth connected surface $\tilde{S}$ in the space
\footnote{  We use slightly different notation than in \cite{23} --- the
present one is more convenient, especially in Section 4. }.
It is called {\it the core} of the domain wall. The transverse width of
such a domain wall is of the order
$M^{-1}$ and energy density is exponentially localised around the core.
The  world-volume $\tilde{\Sigma}$ of the core is a 3-dimensional manifold 
embedded in Minkowski space-time. The core plays an essential
role in the effective membrane models mentioned in the Introduction.

In our approach we use another smooth connected surface $S$ attached
to the domain wall. In general it differs from the core, except at the
initial instant of time when it coincides with the core by assumption.
We shall obtain  Nambu-Goto equation for $S$, so this surface
can be regarded as the Nambu-Goto type relativistic membrane co-moving
with the domain wall. Let us stress that this {\it co-moving membrane}
is merely an auxiliary mathematical notion. It is introduced in order to
define the co-moving coordinate system. The world-volume of $S$ is
denoted by $\Sigma$. We shall parametrise it as follows
\begin{equation}
\Sigma \ni  ( Y^{\mu} ) (u^a) = ( \tau, Y^i( u^a) ).
\end{equation}
We use here the notation $(u^a)_{ a=0, 1, 2} = (\tau, \sigma^1, \sigma^2)$, 
where $\tau$ coincides with the laboratory frame time $x^{0}$, while 
$\sigma^1, \sigma^2$ parametrise the co-moving membrane  $S$ at each 
instant of time. The index $i=1,2,3$ refers to the spatial components
of the four-vector. The points of the co-moving membrane $S$ at the
instant $\tau$ are given by $ (Y^i)(\tau, \sigma^1, \sigma^2) $.
The coordinate system $(\tau,\sigma^1,\sigma^2,\xi)$ co-moving with
the domain wall is defined by the formula
\begin{equation}
x^{\mu} = Y^{\mu}( u^a) + \xi \: n^{\mu}( u^a),
\end{equation}
where $x^{\mu}$ are Cartesian laboratory frame coordinates in Minkowski
space-time, and $(n^{\mu})$ is a normalised space-like four-vector
orthogonal to  $\Sigma$ in the covariant sense,
\[
n_{\mu}( u^a )Y^{\mu}_{,a}( u^a ) = 0,
  \;\;\;\;\; n_{\mu}n^{\mu}=1,
\]
where $Y^{\mu}_{,a}\equiv \partial Y^{\mu}/ \partial u^a$. The three
four-vectors $Y_{,a}$ are tangent to $\Sigma$. The definition
(3) implies that $\xi$ and $u^a$ are Lorentz scalars. In the co-moving
coordinates the co-moving membrane is described by the simple Lorentz
invariant condition $\xi=0$.  For points lying on  $S$
the parameter $\tau$ coincides with the laboratory time $x^{0}$, but for
$\xi \neq 0$  in general $\tau$ is not equal to  $x^0$.

In the co-moving coordinates  Lagrangian (1) has the following form
\begin{equation}
{\cal L} = - \frac{M^4}{32\lambda} \left[ \partial_s \phi
\partial \phi_s + (\phi^2 - 1)^2 + \frac{4}{M^2} G^{ab}
\partial_a \phi \partial_b \phi\right],
\end{equation}
where $\partial _s = \partial /\partial s, \partial_a = \partial /
\partial u^a$, and $G^{ab}$ is given in Appendix A. The field $\Phi$
and the coordinate $\xi$ have been expressed by dimensionless $\phi$
and $s$,
\[
\Phi(x^{\mu}) =  \frac{M}{2\sqrt{\lambda}} \phi(s, u^a),
\;\; \xi = \frac{2}{M} s.
\]

According to  papers \cite{23} the core is shifted with respect to
the co-moving membrane, that is the scalar field has non-zero values on
the membrane. We take this fact as the starting point for the new
derivation of the expansion in the width. The idea consists in
extracting from the scalar field its component living on the co-moving
membrane and treating it separately from the remaining part of the
scalar field.  To achieve this we use the identity
\begin{equation}
\phi(s, u^a) = B(u^a) \psi(s) + \chi(s, u^a),
\end{equation}
where 
\begin{equation}
 B(u^a) \stackrel{df}{=} \phi(0, u^a)
\end{equation}
is the component of the scalar field living on the co-moving membrane, and
\begin{equation}
\chi \stackrel{df}{=} \phi(s, u^a) - B(u^a) \psi(s)
\end{equation}
is the remaining part.
The auxiliary, {\it fixed} function $\psi(s)$ depends on the variable
$s$ only. It is smooth, concentrated around $s = 0$, and
\begin{equation}
\psi(0) = 1.
\end{equation}
It follows that 
\begin{equation}
\chi(0, u^a) = 0.
\end{equation}
We shall see that the best choice for $\psi(s)$ is given by formula
(22) below.

Formulas (5-7) can be regarded as an invertible change of variables
\[ \phi(s, u^a) \rightarrow \left( B(u^a), \chi(s, u^a) \right) \]
in the configuration space of the scalar field. Therefore, it is legitimate
to use formula (5) in Lagrangian (4) and to derive Euler-Lagrange equations
by taking independent variations of $B(u^a)$ and $\chi$. The
variation $\delta \chi$  has to respect the condition (9), hence
\[
\delta \chi(0, u^a) = 0.
\]
Because of this condition,  variation of the action functional
\[ {\cal S} = \frac{2}{M} \int ds d^3u \; \sqrt{-g} h(s, u^a) {\cal L}
\]
with respect to $\chi$ gives Euler-Lagrange equation in the regions
$s < 0$ and $s > 0$. It has the following form
\begin{eqnarray}
\lefteqn{\frac{2}{M^2}  \frac{1}{\sqrt{-g}  }
\partial_a [\sqrt{-g} h G^{ab} \partial_b ( B \psi +
 \chi ) ]   } \\
& & + \frac{1}{2 }  \partial_s [ h \partial_s ( B \psi + \chi)] 
 + h (B \psi + \chi) [1 - (B \psi + \chi)^2] = 0, \nonumber
\end{eqnarray}
where $h,g$ and $G^{ab}$ are given in Appendix A.  At $s = 0$
there is no Euler-Lagrange equation corresponding to the variation
$\delta\chi$. Instead, we have the condition (9). Equation (10)
should be solved in the both regions separately, with (9) regarded as
a part of boundary conditions for $\chi$. To complete the boundary
conditions we also specify the behaviour of $\chi$ for $|\xi|$ much larger
than the characteristic length $1/M$, that is for $|s| \gg 1$. In our model
(1) the expected behaviour of the domain wall field $\Phi$ for 
large $|\xi|$ is given by an exponential
approach to the vacuum values. Therefore, we shall seek a solution such that
 $\chi$  is exponentially close to +1 for $s \gg 1$ , while for $s \ll -1$
it is exponentially close to $-1$.

At this stage of our considerations Eq.(10) should not be extrapolated
to $s = 0$. For example, the l.h.s. of it could  have a $\delta(s)$-type
singularity. It would occur if $\chi$ was smooth for $s>0$ and for $s<0$
but had a spike at $s = 0$.

In addition to Eq.(10) we also have the Euler-Lagrange equation
corresponding to variations of $B(u^a)$. This equation has the
following form
\begin{eqnarray}
\lefteqn{\frac{2}{M^2} \int ds \; \frac{1}{\sqrt{-g} } 
\partial_a \left[\sqrt{-g} h G^{ab} \partial_b ( B \psi +
\chi ) \right]  \psi }  \\
&  & - \frac{1}{2} \int ds \; h \partial_s \psi \partial_s ( B \psi + \chi )
 \nonumber \\
& &  + \int ds \; h \psi ( B \psi +  \chi ) \left[ 1 - ( B \psi + \chi )^2
\right] = 0. \nonumber
\end{eqnarray}
Here and in the following we use $\int ds$ as a shorthand for the definite
integral $\int_{-\infty}^{+\infty} ds$.

Comparing equations (10) and (11) one might think that they are not
independent because it seems that multiplying Eq.(10) by
$ \psi$ and integrating
the result over $s$ with the help of integration by parts we obtain
Eq.(11).  This argument is false, namely it ignores the above
mentioned possibility that  the l.h.s. of Eq.(10) might have the
$\delta$-type singularity at $s = 0$. On the other hand, if 
the singularity at $s=0$ is absent then indeed Eq.(11) does follow from
Eq.(10).

Actually, just because  Eqs.(10) and (11) are independent we can prove
that the solution $\chi$ does not have the spike at $s = 0$.
To this end, we multiply (10) by $\psi$
and integrate over $s$ in the intervals $(\infty, \epsilon], [-\epsilon,
-\infty)$ with a positive $\epsilon$ which approaches zero. In this
manner we avoid the not-excluded-yet singularity at $s = 0$.
The l.h.s. of the resulting formula is compared with the l.h.s.
of Eq.(11):  in the limit $\epsilon \rightarrow 0$ they differ by the term
\[
- \frac{1}{2} \psi h \partial_s \chi |_{s=0+}
+ \frac{1}{2} \psi h \partial_s \chi|_{s=0-},
\]
which has to vanish  because the 
r.h.s.'s of Eqs.(10), (11) vanish. In conclusion,
\[
\lim_{s\rightarrow 0+} \partial_s \chi =
\lim_{s\rightarrow 0-} \partial_s \chi,
\]
that is $\partial_s\chi$ is continuous at $s = 0$. Hence, Eq.(10) and 
Eq.(11) together imply that the spike at $s = 0$ is absent. It follows that 
Eq.(10) is obeyed by $\chi$ also at $s = 0$. To summarize, Eq.(11) and
the statement that the singularity at $s=0$ is absent are equivalent.

Let us now solve Eq.(10) in the leading approximation which is
obtained by putting $1/M = 0$. The equation is then reduced to
\begin{equation}
\frac{1}{2}  \partial^2_s \phi^{(0)} +  \phi^{(0)} [1- (\phi^{(0)})^2] = 0,
\end{equation}
where
\[ \phi ^{(0)} = B^{(0)} \psi + \chi ^{(0)}.  \]
In the $1/M$ expansion this is the only nonlinear equation we have
to solve. Mathematically, it coincides with a well-known equation
for a planar domain wall. It has the following particular solutions
\begin{equation}
\phi^{(0)}_{\pm}(s, u^a) = \pm \tanh( s - b(u^a) ),
\end{equation}
where $b(u^a)$ is an arbitrary function of the indicated variables. In
the following we shall take the $+$ sign  (the $-$ sign corresponds
to an anti-domain wall). The functions $b(u^a)$ can be
removed from the solution (13) by a suitable choice of the co-moving
membrane. The new membrane is obtained by shifting the points of
the worldsheet $\Sigma$ of the original
membrane along the direction normal to $S$ by the distance $b(u^a)$.
Because the normal direction is given locally by the $ n(u^a)$
four-vector, this shift is equivalent to the change $(\xi, u^a)
\rightarrow (\xi', u^a)$, $\xi'=\xi - b(u^a)$ of the co-moving coordinates.
Thus, we may take as the zeroth order domain wall solution in the
co-moving frame
\begin{equation}
\phi^{(0)} = \tanh s, 
\end{equation}
without any loss of generality. This solution together with conditions 
(8), (9) gives 
\begin{equation}
B^{(0)} =0, \;\;\; \chi^{(0)} = \tanh s.
\end{equation}

The solution (14) in the co-moving coordinates does not determine 
the field $\phi$ in the laboratory frame because we do not know yet the 
position of the co-moving membrane with respect to the laboratory frame.
Equations (10), (11) should also yield an equation
for the co-moving membrane, otherwise they would not form the complete
set of evolution equations for the domain wall.
In fact, we shall see that the first order terms in Eq.(10)
imply the Nambu-Goto equation for the membrane.

The expansion in the width in the present formulation has the form
\begin{equation}
\chi(s, u^a) = \tanh s + \frac{1}{M} \chi^{(1)}(s,u^a) +
\frac{1}{M^2} \chi^{(2)}(s,u^a)   + \frac{1}{M^3}  \chi^{(3)}(s,u^a) + ... ,
\end{equation}
\begin{equation}
B(u^a) =  \frac{1}{M} B^{(1)}(u^a) +  \frac{1}{M^2} B^{(2)}(u^a)
  + \frac{1}{M^3}  B^{(3)}(u^a) + ... ,
\end{equation}
where we have taken into account the zeroth order results (15).
The expansion parameter is $1/M$
and not $1/M^2$ because $1/M$ in the first power appears in the $h$ and
$G^{ab}$  functions after passing to the $s$ variable, see formulas in 
Appendix A.  In order to obey the condition (9) and to ensure the proper
asymptotics of $\chi$ at large $|s|$ we assume that for $n \geq 1$
\begin{equation}
\chi^{(n)}(0, u^a) = 0, \;\;\; \lim_{s \rightarrow \pm \infty}
\chi^{(n)} = 0.
\end{equation}
Inserting the perturbative Ansatz (16), (17) in Eqs.(10) and (11),
expanding the l.h.s.'s of them in powers of $1/M$, and equating to zero
coefficients in front of the powers of 1/M we obtain a sequence of 
linear, inhomogeneous equations for $\chi^{(n)}(s, u^a), B^{(n)}(u^a)$ with
$n\geq 1$ . 

The first order terms in Eq.(10) give the following equation
\begin{equation}
 \hat{L} \chi^{(1)} = - B^{(1)} \hat{L} \psi 
+  K_a^a  \partial_s \chi^{(0)},
\end{equation}
where
\[
\hat{L} \stackrel{df}{=} \frac{1}{2} \partial_s^2 + 1 - 3 (\chi^{(0)})^2,
\]
and $\chi^{(0)}$ is given by the second of formulas (15). General 
solution of the equation of the form (19) is presented in Appendix B.

The most important point in our derivation of the expansion in the width
is the observation that the operator $\hat{L}$ has a zero-mode, that is
the normalizable solution 
\[
\psi_0(s) = \frac{1}{\cosh^2s} 
\]
of the homogeneous equation
\begin{equation}
\hat{L}\psi_0 = 0.
\end{equation}
Notice that $\psi_0 = \partial_s \chi^{(0)}$ --- this means that the
zero-mode $\psi_0$ is related to the translational invariance of Eq.(12)
under $s \rightarrow s$ + const.
The presence of the zero-mode implies the consistency condition: we
multiply Eq.(19) by $\psi_0$ and integrate over $s$. With the help
of integration by parts we show that $\int \psi_0 \hat{L} \chi^{(1)}$
and  $\int \psi_0 \hat{L} \psi$ vanish because of (20). Finally, we
obtain the following condition
\[
K^a_a \int ds \; \psi_0(s) \partial_s \chi^{(0)}(s) = 0
\]
which is equivalent to
\begin{equation}
K^a_a = 0.
\end{equation}
This condition coincides with the well-known Nambu-Goto equation. It 
determines the motion of the co-moving membrane, that is the 
functions $Y^i(u^a)$,
i=1,2,3, once initial data are fixed. When we know these functions we
can calculate the extrinsic curvature coefficients $K_{ab}$ and the
metric $g_{ab}$.  Review of properties of relativistic Nambu-Goto
membranes can be found in, e.g., \cite{29}.

Due to Nambu-Goto equation (21) one term on the r.h.s. of Eq.(19)
vanishes. The resulting equation
\[
\hat{L} \chi^{(1)} = -  B^{(1)} \hat{L} \psi
\]
has the following solution obeying the boundary conditions (18)
\[
\chi^{(1)} =  B^{(1)}(u^a) ( \psi_0(s) - \psi(s) ). 
\]
It can be obtained from the formulas given in Appendix B, but there is a
shorter way: notice that $\chi^{(1)} + B^{(1)} \psi $ obeys the
homogeneous equation (20)
which has the general solution $C(u^a)\psi_0(s) + D(u^a) \psi_1(s)$; the
boundary conditions (18) imply that $D=0, C=B^{(1)}$.
Now, we recall that the function $\psi(s)$ is an auxiliary mathematical
object, like the co-moving membrane in our approach, and we may adjust
it in order to simplify formulas.  The obvious choice is
\begin{equation}
\psi(s) = \psi_0(s) = \frac{1}{\cosh^2s},
\end{equation}
because then the first order correction has the simplest possible form
\begin{equation}
\chi^{(1)}(s, u^a) = 0.
\end{equation}
In the remaining part of our paper we will use $\psi$ given by
formula (22). Notice that vanishing $\chi^{(1)}$ does not mean that
the first order correction to the original field $\phi$ also vanishes ---
there is the first order contribution equal to $B^{(1)} \psi_0/M$.
It does not vanish on the co-moving membrane that is at $s = 0.$

Equations (10), (11) in the first order do not give any restriction
on the function $B^{(1)}$. In the next Section we shall see that an
equation for $B^{(1)}$ (Eq.(32) below) follows from the third order 
terms in Eq.(11). This situation is typical for
singular perturbation theories of which the $1/M$ expansion is an example
--- higher order equations imply restrictions (the consistency conditions)
for the lower order contributions \cite{27}.

With the choice (22) for $\psi$, Eq.(11) expanded in the powers of 
$1/M$ gives equations for $B^{(n)}$ coinciding with the
consistency conditions found in \cite{23}. This follows from the fact that
both Eq.(11) and the consistency conditions are obtained by multiplying
expanded with respect to $1/M$ Eq.(10) by the zero-mode $\psi_0$ and
next integrating over $s$. Euler-Lagrange equation (11) can be regarded
as {\it the generating equation} for the consistency conditions.

\section{ The higher order corrections }

At this point we have the complete set of equations determining the
evolution of the domain wall in the $1/M$ expansion. 
Each of Eqs.(10), (11), (21) describes
a different aspect of the dynamics of the curved domain wall.  We shall see
that expanded in the powers of $1/M$ Eq.(10) determines dependence
of $\chi$ on $s$, that is on the distance from the worldsheet
$\Sigma$ of the co-moving membrane along the perpendicular
direction $n(u^a)$ at each point $Y(u^a)$ of the worldsheet.
Because the term $B\psi$ in formula (5) has explicit dependence on $s$,
see formula (22), we may say that Eq.(10) for $\chi$ fixes the
transverse profile of the domain wall.

Equation (11) determines the  $B^{(n)}(u^a)$ functions, which can be regarded
as a (2+1)-dimensional scalar fields defined on the worldsheet
$\Sigma$ and having nontrivial nonlinear dynamics, see
Eqs.(32), (35) below. The extrinsic curvature $K_{ab}$ of $\Sigma$ acts as 
an external source for these fields.  The  fields $B^{(n)}$ can propagate
along $\Sigma$. One may regard this effect as {\it causal} propagation
of deformations which are introduced by the extrinsic curvature.

Finally, Nambu-Goto equation (21) for the co-moving membrane determines
the evolution of the shape of the domain wall.

Equations (21) and (11) are of the evolution type in the $1/M$ expansion 
--- we have to
specify initial data for them, otherwise their solutions are not
unique. Equations for the perturbative contributions $\chi^{(n)}$ are
of different type --- in order to ensure uniqueness of their solutions
it is sufficient to adopt the boundary conditions (18). The initial
data for $B(u^a)$ and $Y^i(u^a)$ follow from initial data for the
original field $\phi$. From such data for $\phi$ we know the
initial position and velocity of the core.  As in papers \cite{23}
we assume that at the initial instant $\tau_0$ the co-moving membrane
and the core have the same position and velocity. Hence,
\[
\mbox{initial data for the membrane = initial data for the core.}
\]
Using formula (5) one can show that then
\begin{equation}
B^{(n)}(\tau_0, \sigma^1, \sigma^2) = 0, \;\;\;
\partial_{\tau} B^{(n)}(\tau_0, \sigma^1, \sigma^2) = 0.
\end{equation}
Our perturbative scheme with the initial conditions (24)
gives {\it the basic curved domain wall}. In Section 5 we
discuss more general solutions obtained by adopting more general
than (24) initial data for the $B^{(n)}$ fields.

In order to find the domain wall solution one should first solve the
collective dynamics, that is to compute evolution of the co-moving
membrane and of the $B$  field. The profile $\chi$ of the domain wall
is found in the next step from formulas (16) above and (27) below. 
In our perturbative
scheme the profile of the domain wall can not be chosen arbitrarily even
at the initial time --- it is fixed uniquely once the initial data
for the membrane and for the $B$ field are given. Evolution of the
core can be determined afterwards, from the explicit expression for
the scalar field $\phi$. This will be done in Section 4.

Now we shall show that order by order in the $1/M$ expansion
Eq.(10) determines the dependence of the field $\chi$ on $s$. $\chi$
will be explicitly expressed by the perturbative contributions to
the $B$ field and by the geometric characteristics $g_{ab}, K_{ab}$ of the
worldsheet of the co-moving membrane. Therefore, the dynamics of the curved
domain wall is reduced to the Nambu-Goto dynamics of the co-moving
membrane and to nonlinear dynamics of the $B^{(n)}$ fields.
A convenient starting point for this calculation is obtained by
rewriting Eq.(10) in a different form. We  substitute
\[ \chi = \chi^{(0)} + \bar{\chi}, \;\;\; h = 1 + \bar{h},\;\;\;
G^{ab} = g^{ab} + \bar{G}^{ab},
\]
where 
\[ \bar{h} = \frac{1}{M^2} h^{(2)} + \frac{1}{M^3} h^{(3)} \]
with
\[
h^{(2)} = -2 s^2 K^b_a K^a_b,  \;\;\; h^{(3)} = -\frac{8}{3} s^3 K^a_b
K^b_c K^c_a,
\]
and 
\[
\bar{G}^{ab} = \frac{1}{M} G^{(1)ab} + \frac{1}{M^2} G^{(2)ab} +
{\cal O}(M^{-3}),
\]
with
\[
G^{(1)ab} = 4s K^{ab},\;\;\; G^{(2)ab} = - 12s^2 K^a_cK^{ca}.
\]
We also use  Eqs.(12), (20) and formulas (15).  In the end Eq.(10) is 
written in the following form
\begin{eqnarray}
\lefteqn{\hat{L} \bar{\chi} = - \frac{2}{M^2} \frac{1}{\sqrt{-g}} 
\partial_a \left[ \sqrt{-g} (1+\bar{h}) (g^{ab}+ \bar{G}^{ab})
\partial_b ( B \psi_0 +
\bar{\chi} ) \right]
- \bar{h} \hat{L}\bar{\chi}  } \\
& & - \frac{1}{2 }   \partial_s \bar{ h} (\psi_0 +
\partial_s \bar{\chi} + B \partial_s \psi_0 )
+ (1 + \bar{h}) ( 3 \chi^{(0)} + B \psi_0 + \bar{\chi} ) (B\psi_0 +
\bar{\chi})^2. \nonumber
\end{eqnarray}
Expanded in the powers of $1/M$ it gives equations of the type considered
in Appendix B, that is
\begin{equation}
\hat{L} \chi^{(n)} = f^{(n)}.
\end{equation}
The source term $ f^{(n)} $ is determined by  the lower order terms
in $\bar{\chi}$ and $B$, namely $f^{(n)}$ contains $\chi^{(k)}$ with
$k\leq n-2$ and $B^{(l)}$ with $l\leq n-1$. Solution of Eq.(26)
is given by the formula
\begin{equation}
\chi^{(n)}(s) = \int dx \; G(s,x) f^{(n)}(x),
\end{equation}
where the Green's function $G(s,x)$ can be found in Appendix B. This
solution obeys the boundary conditions (18).

Let us list $f^{(n)}$ with n=2,3,4:
\begin{equation}
f^{(2)} =  2 s \psi_0 K^a_b K^b_a  + 3 \chi^{(0)} \psi_0^2 (B^{(1)})^2,
\end{equation}
\begin{eqnarray}
\lefteqn{ f^{(3)} = 4 s^2 \psi_0 K^a_bK^b_cK^c_a -2 \psi_0
\Box^{(3)}B^{(1)}   } \\
&  & + f^{(3)}_1(s) K^a_bK^b_a B^{(1)}
+  f^{(3)}_2(s) (B^{(1)})^3 + 6 \chi^{(0)} \psi^2_0 B^{(1)}
B^{(2)}, \nonumber
\end{eqnarray}
\begin{equation}
f^{(4)} = f^{(4)}_1 + f^{(4)}_2 ,
\end{equation}
where
\begin{eqnarray}
\lefteqn{ f^{(4)}_1 = -2 \Box^{(3)}( \chi^{(2)} + \psi_0 B^{(2)})
- 4s\psi_0  \frac{1}{\sqrt{-g} } \partial_a( \sqrt{-g} K^{ab}
\partial_b B^{(1)})   }  \nonumber \\
&  & - \frac{1}{2} \partial_sh^{(3)}\partial_s\psi_0 B^{(1)}
- \frac{1}{2} \partial_sh^{(2)}(\partial_s\psi_0 B^{(2)} + 
\partial_s\chi^{(2)} - h^{(2)}\psi_0) 
 \nonumber \\
& &  + 3   \psi_0^2(B^{(1)})^2  ( \chi^{(2)} + \psi_0 B^{(2)})  + 
3 \chi^{(0)} [ \psi_0^2 (B^{(2)})^2 + (\chi^{(2)})^2 +
2\psi_0 \chi^{(3)} B^{(1)} ], \nonumber
\end{eqnarray}
and
\[
f^{(4)}_2 = 6 \psi_0^2 \chi^{(0)} B^{(1)} B^{(3)}.
\]
Here
\[
\Box^{(3)} \stackrel{df}{=} \frac{1}{\sqrt{-g} } 
\partial_a( \sqrt{-g} g^{ab}\partial_b) 
\]
is the three-dimensional d'Alembertian on the world-volume $\Sigma$
of the co-moving membrane. The functions $f^{(4)}_{1,2}(s)$ can be 
found in Appendix C. The formulas for $f^{(3,4)}$ will become a little 
bit simpler because we shall show that $B^{(2)}$ vanishes.

From formula (27) we obtain the explicit dependence of $\chi^{(n)}$ on
$s$. The integrals over $x$ which appear on the r.h.s. of that formula
after we substitute expressions (28-30) are rather simple.
Most of them can be evaluated analytically and the rest numerically by
a computer algebra system. From the remark at the end of Appendix B and
from formula (28) we see that $\chi^{(2)}$ is an odd function of $s$.
The correction $\chi^{(3)}$ is an even and $\chi^{(4)}$ an odd function
of $s$ provided that $B^{(2)}$ vanishes.

Equations of motion for the (2+1)-dimensional fields $B^{(n)}(u^a)$
are obtained by expanding Eq.(11). It is
convenient to rewrite that equation in the following form 
\begin{eqnarray}
\lefteqn{ \frac{2}{M^2}\int ds\; \psi_0 \frac{1}{\sqrt{-g}} 
\partial_a \left[ \sqrt{-g}(1+\bar{ h}) (g^{ab} + \bar{G}^{ab})
\partial_b ( B \psi_0 +
\bar{\chi}) \right] } \\
& & + \frac{1}{2} \int ds\;  \partial_s \bar{ h}
\left( \psi_0^2 + \bar{\chi} \partial_s \psi_0 +
B \psi_0  \partial_s \psi_0 \right) \nonumber \\
& & - \int ds \;\psi_0 (1 + \bar{h}) 
( 3 \chi^{(0)} + B \psi_0 + \bar{\chi} ) (B\psi_0 + 
\bar{\chi})^2 = 0.  \nonumber
\end{eqnarray}
It has been obtained with the help of integration by parts. We have also 
used Eq.(20). From Eq.(31) we see that the first nontrivial equation 
appears in the order $1/M^3$. It has the form of non-linear, inhomogeneous
(2+1)-dimensional wave equation for $B^{(1)}(u^a)$ regarded
as a field defined on $\Sigma$
\begin{equation}
c_0 \Box^{(3)}B^{(1)}  + c_1  K^a_b K^b_a  B^{(1)} - c_2 (B^{(1)})^3  =
d_0  K^a_b K^b_c K^c_a.
\end{equation}
The constants $c_i, d_0$ are listed in Appendix C.

The equation for  $B^{(2)}$ follows from the $1/M^4$ terms in Eq.(31).
It has the form
\begin{equation}
\Box^{(3)}B^{(2)} + \frac{1}{2}  K^a_bK^b_a B^{(2)} -
\frac{c_3}{c_0} (B^{(1)})^2 B^{(2)} = 0.
\end{equation}
This equation is homogeneous with respect to $B^{(2)}$. Because also the
initial data (24) are homogeneous, $B^{(2)}$ vanishes
\begin{equation}
B^{(2)}(u^a) = 0.
\end{equation}

The $1/M^5$ terms give the equation for $B^{(3)}$. It can be written in 
the form
\begin{equation}
c_0 \Box^{(3)}B^{(3)} + c_1 K^a_bK^b_a  B^{(3)}  -
3 c_2 (B^{(1)})^2 B^{(3)} = j^{(5)},
\end{equation}
where the source term $j^{(5)}$ does not contain $B^{(3)}$. 
It is given by the following formula
\begin{eqnarray}
\lefteqn{ j^{(5)} =  -  \int ds\;  \frac{1}{\sqrt{-g} }
\partial_a[ \sqrt{-g} ( g^{ab}\partial_b\chi^{(3)}\psi_0 + G^{(1)ab}
\partial_b\chi^{(2)}\psi_0 }   \\
& & + h^{(2)} g^{ab}\partial_bB^{(1)}\psi_0^2 +
G^{(2)ab}\partial_bB^{(1)}\psi_0^2) ]      \nonumber \\
& & - \frac{1}{4}\int ds\; \partial_s\psi_0 (\partial_sh^{(3)}\chi^{(2)} +
\partial_sh^{(2)}\chi^{(3)} ) + \frac{1}{2} \int ds\;
\psi_0^4 h^{(2)} (B^{(1)})^3
\nonumber \\
& & + \frac{3}{2} \int ds\; \psi_0 [2 \chi^{(0)} \chi^{(2)} \chi^{(3)} +
\psi_0^2 \chi^{(3)} (B^{(1)})^2 + \psi_0 (\chi^{(2)})^2 B^{(1)}
\nonumber \\ & &
+ \psi_0^2 \chi^{(0)} h^{(3)} (B^{(1)})^2 + 2 \psi_0 h^{(2)} \chi^{(0)}
\chi^{(2)} B^{(1)} ]  \nonumber \\
& & + 3 \int ds\; \psi_0^2 \chi^{(0)} G[f^{(4)}_2] B^{(1)}, \nonumber
\end{eqnarray}
where 
\[
G[f^{(4)}_2](s) = \int dx\;G(s,x) f^{(4)}_2(x).
\]
The function $B^{(2)}$ is not present in formula (36) because we have
taken into account the result (34).

Finally, from Eq.(31) in the $1/M^6$ order we obtain a homogeneous wave
equation for the function  $B^{(4)}$ and therefore, similarly as in the 
case of the function $B^{(2)}$ with Eq.(33), we conclude that
\begin{equation}
B^{(4)}(u^a) = 0.
\end{equation}

Let us summarize our results. We have found that
\begin{eqnarray}
\lefteqn{\phi = \chi^{(0)}(s) + \frac{1}{M}B^{(1)}(u^a)\psi_0(s)
 + \frac{1}{M^2}\chi^{(2)}(s,u^a)  } \\  & &
 + \frac{1}{M^3}B^{(3)}(u^a)\psi_0(s) + \frac{1}{M^3}\chi^{(3)}(s,u^a)
 + \frac{1}{M^4} \chi^{(4)}(s,u^a) + {\cal O}(M^{-5}),   \nonumber
\end{eqnarray}
where  $\chi^{(0)}, \psi_0$ are given by formulas (15), (22), $\chi^{(n)}$
are given by formula (27), and $B^{(1)}, B^{(3)}$ are solutions of
Eqs.(32), (35) with the initial data (24). Before solving Eqs.(32), (35) 
one should first solve Nambu-Goto equation (21) for the co-moving membrane.
This is necessary in order to determine  $g_{ab}$ and $K_{ab}$ which 
appear in these equations, and also
in order to find the explicit form of the transformation (3) 
which relates the co-moving coordinates to the Cartesian laboratory 
coordinates.

\section{Equation of motion for the core}

In the description of the collective dynamics of the domain wall we could
use the core instead of the co-moving membrane. This might even seem a
better choice because the definition of the core as the locus
of zeros of the scalar field $\phi$ is very simple and it directly refers
to the domain wall solution. In this Section we derive the
equation of motion for the core in our approach. We shall see that this
equation is more complicated than the Nambu-Goto equation for the co-moving
membrane and that it involves the $B^{(n)}$ fields.

The worldsheet $\tilde{\Sigma}$ of the core $\tilde{S}$ is related 
to the worldsheet $\Sigma$ of the co-moving membrane by the formula
\begin{equation}
\tilde{\Sigma} \ni \tilde{Y}^{\mu}(u^a) =  
Y^{\mu}(u^a) + \frac{2\bar{s}}{M} n^{\mu}(u^a),
\end{equation}
where $\bar{s}$ is determined from the equation 
\begin{equation}
\phi(\bar{s}, u^a) = 0.
\end{equation}
Taking for $\phi$ the expansion (38) we find that 
\begin{equation}
\bar{s} = - \frac{B^{(1)}}{M} + {\cal O}(M^{-3}).
\end{equation}
We parametrise the core by the same coordinates $(u^a)$ as for the
co-moving membrane. The induced metric tensor and the extrinsic
curvature for $\tilde{\Sigma}$ are defined by the formulas
\begin{equation}
\tilde{g}_{ab} = \tilde{Y}_{,a} \tilde{Y}_{,b}, \;\;\;
\tilde{K}_{ab} = - \tilde{n}_{,a} \tilde{Y}_{,b} =  \tilde{n}
\tilde{Y}_{,ab},
\end{equation}
where the normal to $\tilde{\Sigma}$
\[
\tilde{n}(u^a) = n(u^a) - \frac{2}{M^2} g^{ab} B^{(1)}_{,a} Y_{,b}
+{\cal O}(M^{-4})
\]
has been determined from the condition
\[ 
\tilde{n} \tilde{Y}_{,a} = 0\;\;\;\; \mbox{for} \;\;\;\; a=0,1,2.
\]
We have also used the Nambu-Goto equation $K^a_a =0$.
Simple calculations give
\begin{eqnarray}
&\tilde{g}_{ab} = g_{ab} + \frac{4}{M^2} B^{(1)} K_{ab} + {\cal O}(M^{-4}),
& \\
& \tilde{K}_{ab} = K_{ab} + \frac{2}{M^2}B^{(1)} K_{ac}K^c_b - \frac{2}{M^2}
\nabla_bB^{(1)}_{,a} + {\cal O}(M^{-4}), &
\end{eqnarray}
where 
\[ 
\nabla_bB^{(1)}_{,a} = B^{(1)}_{,ab} - \Gamma^c_{ab} B^{(1)}_{,c}
\]
is the standard covariant derivative of $B^{(1)}_{,a}$ with respect to the
metric $g_{ab}$ on the co-moving membrane.  The relations inverse to (43),
(44) have the form
\begin{eqnarray}
& g_{ab} = \tilde{g}_{ab} - \frac{4}{M^2}B^{(1)}\tilde{K}_{ab}
+ {\cal O}(M^{-4}),  &  \\
& K_{ab} = \tilde{K}_{ab} - \frac{2}{M^2} B^{(1)}\tilde{K}_{ac}
\tilde{K}^c_b + \frac{2}{M^2} \tilde{\nabla}_bB^{(1)}_{,a} +
{\cal O}(M^{-4}), &
\end{eqnarray}
where the tilda over $\nabla$ means that now we use the metric
$\tilde{g}_{ab}$. Using formulas (45), (46) we find that the
equation $K^a_a=0$ for the  co-moving membrane implies
the following equation for the core
\begin{equation}
\tilde{K}^a_a = - \frac{6}{M^2}B^{(1)} \tilde{K}^a_b\tilde{K}^b_a
-  \frac{2}{M^2} \tilde{\Box}^{(3)} B^{(1)} + {\cal O}(M^{-4}).
\end{equation}
Using Eq.(32) we eliminate $\tilde{\Box}^{(3)}B^{(1)}$. Then
\begin{eqnarray}
\lefteqn{ \tilde{K}^a_a = - \frac{2}{M^2}(3-\frac{c_1}{c_0}) B^{(1)}
\tilde{K}^a_b\tilde{K}^b_a -  \frac{2}{M^2}\frac{c_2}{c_0} (B^{(1)})^3
} \\
& & - \frac{2}{M^2} \frac{d_0}{c_0} \tilde{K}^a_b \tilde{K}^b_c
\tilde{K}^c_a + {\cal O}(M^{-4}). \nonumber
\end{eqnarray}
This is the final form of the equation of motion for the core.

The terms on the r.h.s. of Eq.(48) can be regarded as corrections to
the Nambu-Goto equation $\tilde{K}^a_a = 0$. They vanish if
$ \tilde{K}^a_b \tilde{K}^b_c \tilde{K}^c_a = 0$ because then also 
$B^{(1)} = 0$ as it follows from Eq.(32) with the initial conditions (24).
In general, Eq.(48) has to be considered together with Eq.(32)
for $B^{(1)}$ in which we may replace $K_{ab}$ by $\tilde{K}_{ab}$.
It is clear that the preferred by us description of
the collective dynamics in terms of the co-moving membrane of the
Nambu-Goto type instead of the core is much simpler.

In the literature one can find many attempts to represent the collective
dynamics of the domain wall by the core only. In our approach
this would amount to expressing the $B^{(1)}$ field in Eq.(48) by
the extrinsic curvatures. In principle this is possible because
Eq.(32) with the initial data (24) gives a one-to-one relation between
$B^{(1)}$ and $\tilde{K}^a_b\tilde{K}^b_c\tilde{K}^c_a$. However, the
presence on the l.h.s. of Eq.(32) of the operator $\Box^{(3)}$ acting
on $B^{(1)}$ has the consequence that in general $B^{(1)}$ depends on
$\tilde{K}^a_b\tilde{K}^b_c\tilde{K}^c_a$ in a nonlocal manner. Hence,
the resulting self-contained equation for the core will be nonlocal too.

\section{Discussion}

1. The (2+1)-dimensional fields $B^{(n)}(u^a)$ are an important component
in our version of the expansion in the width. The leading term $B^{(1)}/M$
does not vanish if $ K^a_b K^b_c K^c_a  \neq 0$. Using formulas from
the second of papers \cite{26} one can show that  $ K^a_b K^b_c K^c_a  $ 
is proportional to the product of the mean and Gaussian curvatures  
in a local rest frame of an infinitesimal piece of the co-moving membrane.  
It is also shown in that paper that the 
evolution governed by the Nambu-Goto equation is sensitive only to the
mean curvature in the local rest frame. Therefore, the co-moving membrane,
which obeys the Nambu-Goto equation to all orders,  can not account for
effects which are due to the Gaussian curvature.
This is the reason why for the complete local description of the
collective dynamics of the domain wall we need the $B$ field too.

2. There is an important assumption we have tacitly made: that the
derivatives $\chi^{(n)}_{,a}, B^{(n)}_{,a}$ are of the order $1/M^n$.
It is not satisfied, for example, if $\chi$ and $B$ contain modes 
oscillating with a frequency $\sim M$ which give positive powers of
$M$ upon differentiation with respect to $u^a$. If such oscillating
components were present the counting of powers of $1/M$ would no longer
be so straightforward as in Sections 2 and 3. The assumption excludes 
radiation modes as well as massive excitations of
the domain wall. Therefore, the approximate solution we
obtain gives what we may call {\it the basic curved domain wall}.
To obtain more general domain wall solutions one would have to
change appropriately the approximation scheme. Actually, the fact
that such particular radiationless unexcited curved domain wall exists
is a prediction coming from the $1/M$ expansion. The expansion
yields domain walls of concrete transverse profile --- the dependence
on $s$ is explicit in the approximate solution we construct even at
the initial instant of time.  Once we choose the initial position and
velocity of points of the membrane the dependence of the scalar
field on the variable $s$ at the initial time is given by 
formulas (16), (27). This unique profile is characteristic for the 
basic curved domain wall.

3. The approximate domain wall solution we have obtained
can be generalised by relaxing the initial conditions (24) for
the $B^{(n)}$ fields. These homogeneous initial data forbid
appearence of freely propagating along the co-moving membrane
waves of the $B^{(n)}$ fields --- only the $B^{(n)}$ fields
generated by the extrinsic curvature as the source can appear. Suppose
however that we admit such free waves, that is that the initial data for
$B^{(n)}$ are chosen arbitrarily
\footnote{ Of course, we have to
respect the usual limitations
imposed by consistency of the perturbative approach:
\[ \frac{B^{(n)}}{M^n} \ll 1, \;\;  \frac{B^{(n+1)}}{M} \ll B^{(n)}, \;\;
\partial_a \frac{B^{(n)}}{M^n} \ll M. \]
These bounds ensure that order of magnitude of the perturbative
contributions, in principle given by a power of $K_{ab}/M$, is not changed
because of too large values of the $B^{(n)}$ fields at the initial
time or later.}. 
Now $B^{(2)}$ and $B^{(4)}$ do not have to vanish.
Equations (32) and (33) remain unchanged, but in Eq.(35) new terms
containing $B^{(2)}$ will appear.  Let us also choose certain initial data
for Nambu-Goto equation (21) for the co-moving membrane. Then, 
substitution of the corresponding  unique solutions of 
Eqs.(21), (32), (33) and of (the modified) Eq.(35) in formula (38)
gives an approximate solution
$\phi$ of the domain wall type. These more general
solutions are not excluded by the assumption formulated in the preceding
paragraph because the characteristic frequency of the $B^{(n)}$ fields 
is not related to the mass parameter M. From Eqs.(32), (33), (35) we see 
that the $B^{(n)}$ fields have the effective 
$\mbox{mass}^2$ $\sim K^a_bK^b_a$.

Actually, a simple particular case of solutions analogous to the generalised
ones we are now discussing has been reported in paper \cite{30}. In
that paper certain small amplitude excitations of a straightlinear
vortex in the Abelian Higgs model have been considered. Such excitations
are represented by fields defined on the axis of the straightlinear
vortex. They have been called in \cite{30} the zero-mode fields.
They obey a linear, massless wave equation.  It turns out that  our
$B^{(1)}$ field gives the domain wall counterpart of the zero-mode
field of Ref.\cite{30} if we take a planar domain wall for which
$K_{ab} = 0$, and if we assume  that the  $B^{(1)}$ field is so small
that one may neglect the nonlinear term on the l.h.s. of Eq.(32).

\section{Acknowledgements}
The ideas  presented in our paper were conceived and partially developed 
during our visits to the Niels Bohr Institute. We wish to thank the 
High Energy Group from that Institute for hospitality and support.

\section{Appendix A. The co-moving coordinates}

The extrinsic curvature coefficients $K_{ab}$ and induced 
metrics $g_{ab}$ on $\Sigma$ are defined by the following formulas:
\[
K_{ab} \stackrel{df}{=} n_{\mu} Y^{\mu}_{,ab}, \;\;\;\;
  g_{ab} \stackrel{df}{=} Y^{\mu}_{,a} Y_{\mu,b},
\]
where $a,b=0,1,2$. The covariant metric tensor in the new coordinates
has the following form 
\[
[G_{\alpha\beta}] = \left[ \begin{array}{lr} G_{ab} & 0 \\
0 & 1 \end{array} \right], 
\]
where $\alpha,\beta=0,1,2,3$;\  $\;\;\;\alpha=3$ corresponds to the $\xi$
coordinate; and
\[
G_{ab}=N_{ac} g^{cd} N_{db},\;\;\; 
N_{ac} \stackrel{df}{=} g_{ac} - \xi K_{ac}.
\]
Thus, $G_{\xi\xi}=1, \; G_{\xi a}=0$.
Straightforward computation gives  
\[
\sqrt{-G} = \sqrt{-g} \; h(\xi, u^a),
\]
where as usual  $g\stackrel{df}{=} det[g_{ab}]$, $\;\;$
$G\stackrel{df}{=} det[G_{\alpha\beta} ]$, and
\[
h(\xi, u^a) =
1 - \xi K^{a}_{a} +\frac{1}{2} \xi^2 (K^{a}_{a} K^{b}_{b} - K^{b}_{a}
K^{a}_{b}) -\frac{1}{3} \xi^3  K^{a}_{b} K^{b}_{c} K^{c}_{a}.
\]
Also $g$ can depend on $ u^a $.
For raising and lowering the Latin indices of the extrinsic 
curvature coefficients we use the induced metric tensors 
$g^{ab},\;\; g_{ab}$. 

The inverse metric tensor $G^{\alpha \beta}$ is given by the formula
\[
[G^{\alpha \beta}] = \left[ \begin{array}{lr} G^{ab} &  0 \\
0 &   1 \end{array} \right],
\]
where
\[
G^{ab}= (N^{-1})^{ac} g_{cd} (N^{-1})^{db}.
\]
Explicit formula for $(N^{-1})^{ac}$ has the following form:
\[
(N^{-1})^{ac} = \frac{1}{h} \left\{ g^{ac} [1- \xi K^{b}_{b} +\frac{1}{2}
\xi^2 ( K^{b}_{b} K^{d}_{d} - K^{d}_{b} K^{b}_{d} )] \right.
\]
\[
\;\;\;\;\; +   \left. \xi (1-\xi K^{b}_{b}) K^{ac} + \xi^2
K^{a}_{d} K^{dc} \right\}. 
\]
This is just the matrix inverse to $[N_{ab}]$. It has the upper indices
by definition.

In general, the coordinates $( u^a, \xi)$ are defined locally,  in a 
vicinity of the world-volume $\Sigma$ of the co-moving membrane.
Roughly speaking, the allowed range of the $\xi$ coordinate is determined
by the smaller of the two main curvature radia of the membrane
in a local rest frame. We assume that this curvature radius is
sufficiently large so that on the
outside of the region of validity of the co-moving coordinates there are 
only exponential tails of the domain wall, that is the field $\phi$ is
exponentially close to one of the two vacuum solutions.
More detailed discussion of the region of validity of the
co-moving coordinates can be found in the first of papers \cite{23}.

\section{ Appendix B. Solutions of equation $\hat{L} \chi = f$}

Let us consider the equation 
\[
\frac{1}{2}  \partial^2_s \chi  +  \left( 1- 3 \tanh^2 s\right) \chi = f(s).
\]
The presence of the zero mode implies that it has solutions
only if $f(s)$ obeys the condition
\[
\int ds \; \psi_0(s) f(s) = 0,
\]
where $\int ds$ is our shorthand for $\int_{-\infty}^{+\infty}$.
General solution of our equation can be obtained with the help of 
the standard procedure 
consisting of finding two independent solutions of the corresponding
homogeneous equation and constructing a suitable Green's function,
see, e.g., \cite{31}. The two linearly independent solutions
of the homogeneous equation have the form
\[
\psi_{0}(s) = \frac{1}{\cosh^{2}s},
\]
\[
\psi_{1}(s) = \frac{1}{8}\sinh(2 s) + \frac{3}{8} \tanh s +
\frac{3}{8}  \frac{s}{\cosh^{2}s}.
\]
As the Green's function we take 
\[
G(s,x) = 2 \left[\psi_{1}(s)  \psi_{0}(x)
-   \psi_{0}(s)   \psi_{1}(x) \right] \Theta(s-x) +
2 \psi_0(s) \psi_1(x) \Theta(-x),
\]
where $\Theta$ is the step function. This Green's function vanishes 
at $s = 0$. The general solution of our equation  has the form
\[
\chi  = \alpha \psi_{0}   + \beta \psi_{1}
+ \int dx G(s, x) f(x),
\]
where $\alpha, \beta$ do not depend on $s$. If we require that $\chi$ obeys 
the conditions (18) we have to put $\alpha = \beta = 0.$ In this case,
using the given above formula for the Green's function we can write the
solution $\chi$ in the following form
\[
\chi(s) = 2 \psi_1(s) \int_{-\infty}^s dx \; \psi_0(x) f(x) 
- 2 \psi_0(s) \int_0^s dx \; \psi_1(x) f(x).
\]
This specific formula for $\chi(s)$ is used in Section 3. One can easily
check that if $f$ is even (odd) function then $ \chi$ is even (odd) 
function too.

\section{Appendix C. The list of functions and constants}

It is convenient to introduce the notation
\[
G[f](s) = \int\;dx G(s,x) f(x).
\]
The functions appearing in formula (29) for $f^{(3)}$ have the form
\[
f^{(3)}_1(s) = 2 s \partial_s \psi_0 + 12 \chi^{(0)} \psi_0
G[x\psi_0],
\]
\[
f^{(3)}_2(s) = \psi_0^3 + 18 \chi^{(0)}\psi_0
G[\chi^{(0)} \psi_0^2].
\]
The constants present in Eqs.(32), (33), (35) are given by the following
formulas:
\[
d_0 = 2 \int ds \; s^2 \psi_0^2,
\;\;\;   c_0 = \int ds \;  \psi^2_0,
\]
\[
c_1 = - \frac{1}{2} \int ds \; \psi_0 f^{(3)}_1,
\]
\[
c_2 =  \frac{1}{2} \int ds \;  \psi_0 f^{(3)}_2,
\]
\[
c_3 = \frac{3}{2} \int ds\; \psi_0^4 + 18 \int ds\; \psi^2_0 \chi^{(0)}
G[\chi^{(0)}\psi_0^2].
\]

\end{document}